\documentclass[a4paper,11pt]{article}
\usepackage{jheppub} 
\pdfoutput=1
\usepackage[utf8]{inputenc}
\usepackage{amsmath,amsthm,amssymb,amsfonts,mathtools,bm,bbm,array,float,mathdots,dsfont,mathrsfs}
\usepackage{caption,subcaption,multirow,longtable,lscape,wrapfig}
\usepackage[T1]{fontenc} 
\usepackage{ctable}
\usepackage{notoccite,comment}
\usepackage[toc,page]{appendix}
\usepackage[utf8]{inputenc}
\usepackage[autostyle=true]{csquotes}
\usepackage{tabularx}

\numberwithin{equation}{section}
\numberwithin{figure}{section}
\numberwithin{table}{section}

\begin{document}

\title{Calabi-Yau Links and Machine Learning}
\author[a]{Edward Hirst}
\affiliation[a]{Centre for Theoretical Physics, Queen Mary University of London, E1 4NS, UK}
\emailAdd{e.hirst@qmul.ac.uk}

\preprint{\begin{flushright}
QMUL-PH-24-01
\end{flushright}}
\abstract{Calabi-Yau links are specific $S^1$-fibrations over Calabi-Yau manifolds, when the link is 7-dimensional they exhibit both Sasakian and G2 structures. In this invited contribution to the DANGER proceedings, previous work exhaustively computing Calabi-Yau links and selected topological properties is summarised. Machine learning of these properties inspires new conjectures about their computation, as well as the respective Gr\"obner bases.
}

\maketitle

\section{Introduction}\label{sec:intro}
Since their identification as candidate compactification spaces in superstring theory \cite{Candelas:1985en}, Calabi-Yau manifolds have been a particularly popular area of productivity on the geometric side of theoretical physics.
As the topology of the Calabi-Yau manifold used in compactification determines properties of the subsequent effective field theory, the search for manifolds which reproduce the observed universe has been a central goal in the field.

However, the landscape of Calabi-Yau manifolds is colossal, expected to exceed $10^{10000}$ \cite{Altman:2018zlc}, cementing this problem firmly within the realm of big data.
Since there are too many constructions to analyse directly, statistical methods provide a means for feasibly extracting meaningful analysis, of which methods of machine learning are becoming increasingly prominent \cite{Bao:2022rup,He:2023csq}.

Calabi-Yau manifolds may be constructed through a variety of methods, where the landscape size is dominated by the toric variety construction \cite{Batyrev:1993oya,Kreuzer:1995cd,Bao:2021ofk,Berglund:2021ztg,Berglund:2023ztk}, there are notable popular alternative constructions in terms of complete intersections \cite{Candelas:1987kf,Gagnon:1994ek,Gray_2013,Alawadhi:2023gxa,He:2017aed}, as well as the perhaps the most fundamental construction in terms of weighted projective spaces $\mathbb{P}^n_\textbf{w}$ \cite{CANDELAS1990383,Lynker_1999,Berman:2021mcw,Hirst:2023kdl,Ashmore:2023ajy}.

Extensions of the original superstring theories include theories of higher dimension, such as M-theory, which is manifestly 11-dimensional and requires compactification with a 7-dimensional manifold \cite{SchettiniGherardini:2021iox,Gherardini:2023uyx,Berman:2022dpj}, typically G2 manifolds \cite{Awada:1982pk,Duff:2002rw,House:2004pm}.
This has motivated the investigation into physical applications of exceptional geometries with G2 holonomy, as well as the more general G2-structure \cite{delaOssa:2014lma,delaOssa:2017pqy}.

This is where the motivation for links arises, 7-dimensional Calabi-Yau links have a natural G2-structure, which were shown to solve the heterotic system as compactification spaces in \cite{Calvo-Andrade:2016fti,Lotay2023}.
As well as G2-structure, these links also have a Sasakian structure, which is exciting in its own right for applications in physics \cite{Fabbri:1999hw}.
Additionally, these links are already co-closed such that they may serve as useful starting points for flows to manifolds with full G2 holonomy \cite{Karigiannis_2008,Lotay2022,fadel2023flows}.

In this summary, work from \cite{Aggarwal:2023swe} is reviewed which constructs an exhaustive database\footnote{Data and code available at: \href{https://github.com/TomasSilva/MLcCY7.git}{\texttt{https://github.com/TomasSilva/MLcCY7.git}}.} of Calabi-Yau links from Calabi-Yau 3-folds, defined as hypersurfaces in $\mathbb{P}^4_\textbf{w}$ spaces, more thoroughly introduced in §\ref{sec:data}.
Selected topological properties including some Sasakian Hodge numbers and the Crowley-Nordstr\"om $\nu$ invariant \cite{CNInvariant} are calculated for these links.
Then supervised machine learning methods, discussed in §\ref{sec:ml}, indicate a deeper connection between the links and the weight systems used in their construction.
This connection is further established through the raising of novel conjectures as elucidated in §\ref{sec:conj}.

\section{Link Generation}\label{sec:data}
\begin{wrapfigure}[13]{r}{0.35\textwidth}
    \vspace{-20pt}
    \begin{center}
    \includegraphics[width=0.8\linewidth]{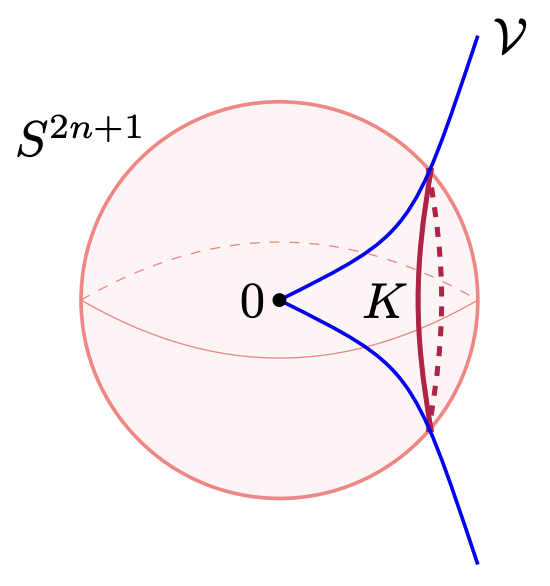} 
    \caption{Abstract diagrammatic for the link construction \cite{Calvo-Andrade2020}.}
    \label{fig:linkdiagram}
    \end{center}
\end{wrapfigure}

A complex variety $\mathcal{V} \subset \mathbb{C}^{n+1}$ with an isolated singularity at the origin can be shown to transversally intersect a sufficiently small sphere $S_\varepsilon^{2n+1}$ centred at the origin \cite{Milnor1969,MILNOR1970385}.
Reducing to the manifold determined by that intersection defines a general link: $K \vcentcolon = \mathcal{V} \cap S_\varepsilon^{2n+1}$; visually represented in Figure \ref{fig:linkdiagram}.
These links are $(n-2)$-connected smooth compact manifolds, and under the scenario that the complex variety is defined by a homogeneous polynomial $f$, the link manifold may be redescribed as the total space of a Hopf $S^1$-bundle over a projective $(n-1)$-manifold defined by the same $f$ within an appropriate weighted projective space $\mathbb{P}_\textbf{w}^n$ \cite{Calvo-Andrade:2016fti}. 

This $S^1$-bundle manifestly produces a global angular contact form $\theta$, with respective dual Reeb vector $\xi$ defining a characteristic codimension-1 foilation with a `transverse-K\"ahler' structure \cite{Portilla2023,Boyer2008}.
These manifolds in general are named Calabi-Yau links, and are built from Calabi-Yau $(n-1)$-folds constructed within $\mathbb{P}_\textbf{w}^{n}$ spaces.

If the link dimension is set to be 7 (via $n=4$) they have a further G2-structure (characterised by a non-degenerate 3-form $\varphi$), as defined by
\begin{equation}
    \varphi \vcentcolon = \theta \wedge \omega + \text{Im} \Omega\;,\qquad 
    \psi = *\varphi \vcentcolon = \frac{1}{2} \omega \wedge \omega + \theta \wedge \text{Re} \Omega \;,
\end{equation}
for 2-form $\omega = d\theta$, and Calabi-Yau 3-fold holomorphic volume (3,0)-form $\Omega$. This G2-structure is naturally coclosed $d\psi = 0$. 

In building a database of Calabi-Yau links with these structures, the finite database of 7555 weight systems $(w_0,w_1,w_2,w_3,w_4)$ defining $\mathbb{P}_\textbf{w}^4$ spaces which exhibit (classes of) Calabi-Yau 3-folds as hypersurfaces as non-singular homogeneous polynomial loci with degree $= \sum_i w_i$ was considered.
Each $\mathbb{P}_\textbf{w}^4$ is defined by the identification
\begin{equation}
    (z_0,z_1,z_2,z_3,z_4) \sim (\lambda^{w_0}z_0,\lambda^{w_1}z_1,\lambda^{w_2}z_2,\lambda^{w_3}z_3,\lambda^{w_4}z_4)\;,
\end{equation}
$\forall \lambda \in \mathbb{C}^\ast$, over the $(\mathbb{C}^\ast)^5$ space spanned by $z_i$, for positive integer weights $w_i$.
For each weight system, one polynomial was selected, checking it to be non-singular away from the isolated singularity at the origin (removed prior to the $\mathbb{P}_\textbf{w}^4$ identification).
For each of these generated links, topological properties associated to their Sasakian and G2 structures were computed, as detailed in the subsequent section.

\paragraph{Example:} As a running example of the generated geometries, consider the weight system $(22,29,49,50,75)$, with $\sum_i w_i = 225$.
The relevant degree $d = 225$ monomial basis consists of the 7 terms: $\{z_0^8z_2,\; z_0^4z_1^3z_2,\; z_0z_1^7,\; z_0z_1z_2z_3z_4,\; z_1z_2^4, \;
z_3^3z_4,\; z_4^3\}$, hence the general Calabi-Yau polynomial equation is
\begin{equation}\label{eq:example}
        0 =  a_1z_0^8z_2 + a_2z_0^4z_1^3z_2 + a_3z_0z_1^7 + a_4z_0z_1z_2z_3z_4 + a_5z_1z_2^4 + a_6z_3^3z_4 + a_7z_4^3\;,
\end{equation}
for complex coefficients $a_i$, which we set all to $1$ ($a_i = 1 \ \forall i$) when selecting the Calabi-Yau 3-fold used to generate the respective link.

\subsection{Topological Invariants}\label{sec:topology}
\begin{figure}[!t]
    \centering
    \begin{subfigure}{0.47\textwidth}
        \centering
        \includegraphics[width=0.98\textwidth]{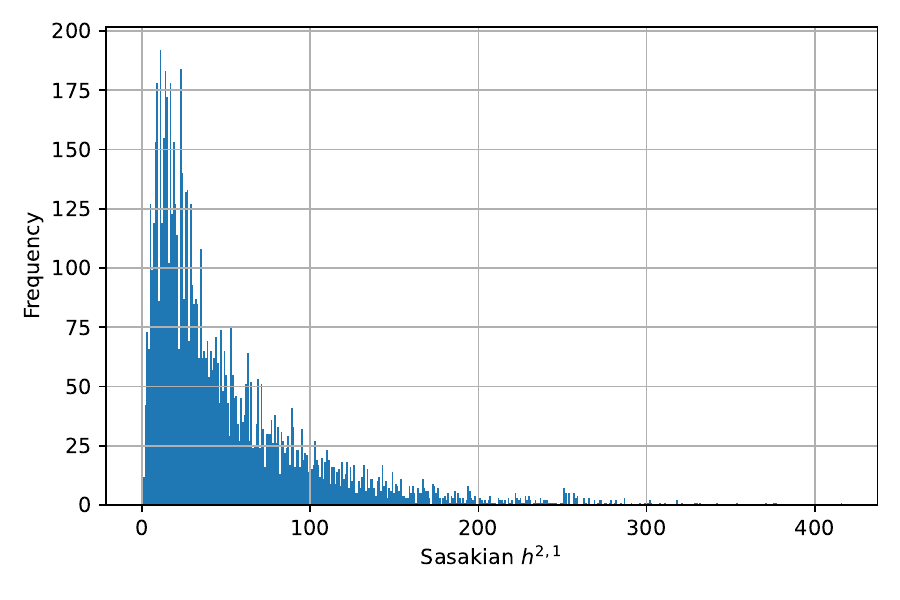}
        \caption{}\label{fig:sh21}
    \end{subfigure} 
    \begin{subfigure}{0.47\textwidth}
        \centering
        \includegraphics[width=0.98\textwidth]{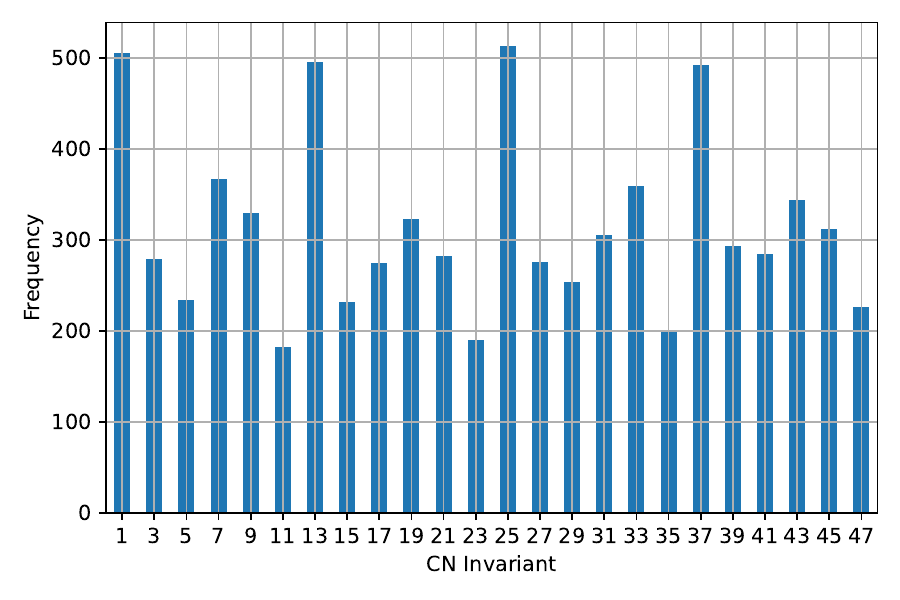}
        \caption{}\label{fig:cni}
    \end{subfigure} 
    \caption{Distributions of the topological invariants (a) $h_S^{2,1}$ and (b) $\nu$ for the database of 7555 Calabi-Yau links generated.}\label{fig:invariants}
\end{figure}

Analysing the topology of the generated database of Calabi-Yau links, work in \cite{Itoh2004} demonstrated that specific Hodge numbers of the transverse K\"ahler structure could be computed as functions of certain linear subspaces of the Milnor algebra $\mathbb{M}_f \vcentcolon = \mathbb{C} [z_i]/(\partial f/\partial z_i)$.
Hence by computing the Gr\"obner basis of this Milnor algebra, via the respective Jacobi ideal for the Calabi-Yau 3-folds defining polynomial $f$, Hodge numbers with $p+q=3$ may be computed as
\begin{equation}\label{eq:shodge}
    h^{p,q}(K) = \dim_\mathbb{C} (\mathbb{M}_f)_\ell\;,
\end{equation}
for $\ell=pd$, enumerating the Milnor algebra subspace of degree $\ell$ elements. 
Therefore for each of the 7555 links in the generated database the respective Gr\"obner basis was computed, and these Sasakian (i.e. transverse K\"ahler) Hodge numbers $\{h^{3,0},h^{2,1}\}$ calculated.
Across the entire database each $h^{3,0}$ number took the value of 1, interestingly matching all the constituent Calabi-Yau manifolds used in their construction (which have a unique holomorphic top form as a defining feature \cite{He:2018jtw,Bao:2020sqg}).
Conversely the $h^{2,1}$ values varied across the database, with distribution shown in Figure \ref{fig:sh21}.
Our running example in \eqref{eq:example} had values $\{h^{3,0},h^{2,1}\} = \{1,2\}$.

In addition to analysis of the link's Sasakian topology, the G2-structure of the links inspires computation of further topological properties, and the one of focus in this work was the Crowley-Nordstr\"om $\nu$ invariant \cite{CNInvariant}.
A $\mathbb{Z}_{48}$-valued homotopy invariant, computed via considering a compact coboundary Spin(7)-structure manifold $(W_8,\Psi)$, such that $K = \partial W$ \& $\Psi |_K = \varphi$; the $\nu$ invariant is then traditionally defined as
\begin{equation}
    \nu(\varphi) := \chi(W) - 3\sigma(W) \quad \text{mod} \ 48
\end{equation}
for Euler characteristic $\chi$ and signature $\sigma$. 
However, work in \cite{CM_1977__34_2_211_0,Calvo-Andrade2020}, demonstrated that for the Calabi-Yau link construction the $\nu$ invariant values could also be computed as a different function of the same Milnor algebra Gr\"obner basis used in \eqref{eq:shodge}.
These were hence also computed for each of the 7555 links considered in the database, producing a distribution as shown in Figure \ref{fig:cni}.

Firstly, it is satisfying that the $\nu$ invariant computations for this new Calabi-Yau link database corroborates \cite[Proposition 3.2]{Calvo-Andrade2020} -- that all links have $\nu$ values which are odd.
Secondly, new values of the invariants for this construction were observed for the first time, notably Calabi-Yau links with $\nu =$ 27 or 35, of which our database exhibited hundreds of examples -- including our running example of \eqref{eq:example} with $\nu = 27$. 
Truly establishing the generation of new topologies.

\section{Learning Topology}\label{sec:ml}
Motivated by the successes in previous works \cite{He:2017aed,Berman:2021mcw,Hirst:2023kdl} using supervised machine learning methods to predict Calabi-Yau Hodge numbers from the ambient $\mathbb{P}_\textbf{w}^n$ weights alone, similar architectures were employed to predict the computed link topological properties from the same weights now used in the link construction.

However, where in the Calabi-Yau case there is a known explicit formula mapping the weights to the respective Calabi-Yau Hodge numbers (for detailed explanation see \cite{Hirst:2023kdl} §5), for the Sasakian Hodge number case and for the $\nu$ invariant, there is a priori no assumed direct connection bypassing the information of \textit{both} the Calabi-Yau polynomial choice, and the particularly expensive computation of the associated Gr\"obner basis.
Initially focus is put on the prediction of $h_S^{2,1}$, as more directly inspired by previous work predicting Hodge numbers, and avoiding trivial learning of the $h_S^{3,0}=1$ values for all links.

Supervised learning is a subfield of machine learning, spanning the range of function fitting techniques trained on pairs of input and output data.
The dataset is split into train and test subsets, where throughout the training procedure an optimiser method adjusts the architecture's parameters across batches of the train dataset to minimise some loss function of the true output and architecture predicted output for each batch input.
After epochs of the training process on the training dataset the final trained architecture is evaluated on its predictions for the independent test dataset's inputs (which it has not seen until training has finished), allowing an evaluation of performance.
The process of training and testing may be repeated on multiple partitions of the original dataset on independent but identically designed architectures to provide a means of averaging and calculating confidence of the learning measures (this is cross-validation).

The architectures used in the work reviewed here were simple feed-forward regressor neural networks \cite{anderson1995introduction}, they had layer sizes of (16, 32, 16), used ReLU activation, and were trained with the Adam optimiser on a mean-squared-error (MSE) loss, then also evaluated with the $R^2$ measure.
The loss function and $R^2$ measure are respectively defined:
\begin{alignat}{3}
    &\text{MSE} && = \frac{1}{N} \sum_{i=1}^{N} (y_{i} - \hat{y}_{i})^2 &&\in [\textbf{0},\infty)\;,\\
    &R^{2} && = 1 - \frac{\sum_{i=1}^{N}(y_{i}-\hat{y}_{i})^{2}}{\sum_{i=1}^{N}(y_{i}-\bar{y})^{2}}
    &&\in (-\infty,\textbf{1}]\;,
\end{alignat}
for true output $y_i$ with mean $\overline{y}$, architecture predicted output $\hat{y}_i$, and subdataset size $N$.
The bold values indicate how each measure evaluates for perfect learning in its range.

The neural network functions were hence attempting to approximate a hypothetical function: $NN(w_i) = h_S^{2,1}$;
results for this learning are displayed in Table \ref{tab:supervised_results}.
In addition to the use of these prototypical neural network architectures, another more interpretable supervised learning method was used: symbolic regression. 

In symbolic regression, a basis of functions (here simply $\{+,-,*,/\}$) is randomly sampled to produce a population of expressions (represented as trees).
These expressions are evaluated on the training dataset, and used to assign competency scores for how well they predict the true outputs from the inputs, as well as a parsimony factor that rewards simplicity in the expression.
The best performing expressions in the population are selected for breeding and mutation to produce the next generation of expressions, and this genetic algorithm process is iterated until a prespecified level of convergence where some expressions in the population well predict the outputs \cite{cranmer2023interpretable}.
Alike the neural networks, this method was also used to predict $h_S^{2,1}$ from the weights $(w_i)$ alone, with performance shown complementarily in Table \ref{tab:supervised_results}.

\begin{table}[!t]
\centering
\begin{tabular}{|c|c|}
\hline
\begin{tabular}[c]{@{}c@{}}Supervised\\ Method\end{tabular} & $h_S^{2,1}$ Prediction \\ \hline
NN $R^2$ & 0.969 $\pm$ 0.003 \\ \hline
SR $R^2$ & 0.99 \\ \hline
\end{tabular}
\caption{Supervised learning results for prediction of the Calabi-Yau links' Sasakian Hodge number $h_S^{2,1}$ from the input Calabi-Yau ambient $\mathbb{P}_\textbf{w}^4$ space's defining weights $(w_0,w_1,w_2,w_3,w_4)$. The NN row details the mean neural network performance with standard error averaged over the 5 cross-validation runs, whilst the SR row states the final symbolic regression performance of the best expression.}
\label{tab:supervised_results}
\end{table}

Both methods perform surprisingly well, raising support for the existence of a currently unknown direct function from weight system to Calabi-Yau link Sasakian Hodge numbers.
The symbolic regression method is particularly useful since the final trained function is more interpretable, and the best performing equation of the provided basis took the form:
\begin{align}
    \label{eq:SymReg_eq}
    h_S^{2,1}(w_0, \dots, w_4)= \frac{14.91 w_{1} \left(w_{0} w_{4} + w_{3} \left(w_{0} + w_{3}\right)\right)
    + 10.02 w_{2} w_{3} \left(w_{0} + w_{4} + 0.77\right)}{w_{0} w_{1} w_{2} w_{3}}\;,
\end{align}
from which one can start to extract mathematical insight, ideally guiding future work towards uncovering the true functional form.
Some initial observations of these results show a unique denominator which omits the largest weight, as well as all but one numerator terms involving 3 weights.

Conversely for the prediction of the $\nu$ invariants, initial experimentation unfortunately showed no fruitful results, implying any direct formula (if it were to even exist) is too complicated to be approximated by the class of functions expressible with simple, small, and finite neural networks.
However, despite no underlying function being approximated, some experimentation with the generation did hint at the existence of some equivalence independent of the polynomial choice, as discussed in §\ref{sec:conj}.

\paragraph{Example:} For our running example of \eqref{eq:example}, the true value of $h_S^{2,1} = 2$.
Our trained architectures predicting this value from the weight system input give values NN: $2.16$ and SR: $1.59$, which when rounded to the nearest integer both give correct predictions of 2. 

\subsection{Compute Optimisation via Machine Learning}\label{sec:grob}
As previously mentioned, the computational bottleneck of the invariant calculations was the generation of an appropriate Gr\"obner basis for each link's Milnor algebra.
Across the database, some of these links could have a basis generated in the order of minutes, whilst others took the order of months.
A priori there was no intuition on what properties of the weight system or polynomial could indicate the Gr\"obner basis computation would be substantially longer, where naive ideas about monomial basis length for the polynomial, or sum of the weights, did not have substantially significant correlations.
Because of this high-performance computing (HPC) cluster job submissions were largely inefficient, with many submitted jobs failing due to limited RAM or wall-time allocations, and conversely many jobs being allocated unneeded memory.

Motivated by strong neural network performances predicting $h_S^{2,1}$ from the input weight system, and with the hindsight of the importance of an appropriate measure to sort the weight systems such that HPC resources can be allocated efficiently for computation, prediction of the Gr\"obner basis length was also investigated.
Work in \cite{jamshidi2023predicting} attempted this for more general yet shorter ideals, reaching performance scores of $R^2 \sim 0.4$.
Here, for the the ideals arising from the weight systems under consideration, the scores were significantly higher: $R^2 = 0.964 \pm 0.002$.

These results are especially exciting, suggesting scope for practical prediction of Gr\"obner basis properties in the wider range of problems their efficient calculation is required.

\paragraph{Example:} For our running example of \eqref{eq:example}, the true length of the Gr\"obner basis for the sorted weight system in the order given is of $209$.
The trained NN architecture predicting this value from the weight system input predicts $225.31$, which when rounded to the nearest integer gives 225. 
This prediction is 16 away form the true value, at an 8\% error, providing a good practical estimate.

\section{Novel Conjectures}\label{sec:conj}
Experimentation computing the considered invariants for a variety of polynomials, with the correct singularity structure and defined from the same weight system, showed the same invariant values each time.
This was tested for hundreds of example polynomials used to build links, and for hundreds (of the quicker to compute) weight systems.

Stimulated by these results, the work summarised here conjectured extension of the $R$-equivalence proved in \cite{ahmed2012}, such that any polynomials with the same weight system would have isomorphic linear subspaces in their respective Milnor algebras. 
This was designated `weak $R$-equivalence', and where the $R$-equivalence of \cite{ahmed2012} states that if two weighted homogeneous polynomials on $\mathbb{C}^{n+1}$ of the same degree have isomorphic Jacobi ideals then a diffeomorphism exists between them, for weak $R$-equivalence in the cases where the Jacobi ideals are \textit{not} isomorphic then the $\ell$-degree linear subspaces of their Milnor algebras are isomorphic, for each $\ell$ given that $p+q=n-1$.
In a trivial generalisation the weight system may also be permuted between the polynomials for weak $R$-equivalence, which although does impact the Gr\"obner basis calculation does not change the final invariant values.
A statement of this conjecture presented in an analogous form to \cite{Aggarwal:2023swe} is

\medskip
\noindent \textbf{Conjecture 1}: \textit{Any two weighted homogeneous polynomials of degree $d$ on $\mathbb{C}^n$ with the same weight vector $\underline{w}$ are `weakly R-equivalent', such that the respective $\ell$-degree linear subspaces of their Milnor algebras are isomorphic, for each $\ell = (p+1)d-\sum_i w_i$}. 
\medskip

\begin{figure}[!t]
    \centering
    \includegraphics[width=0.6\textwidth]{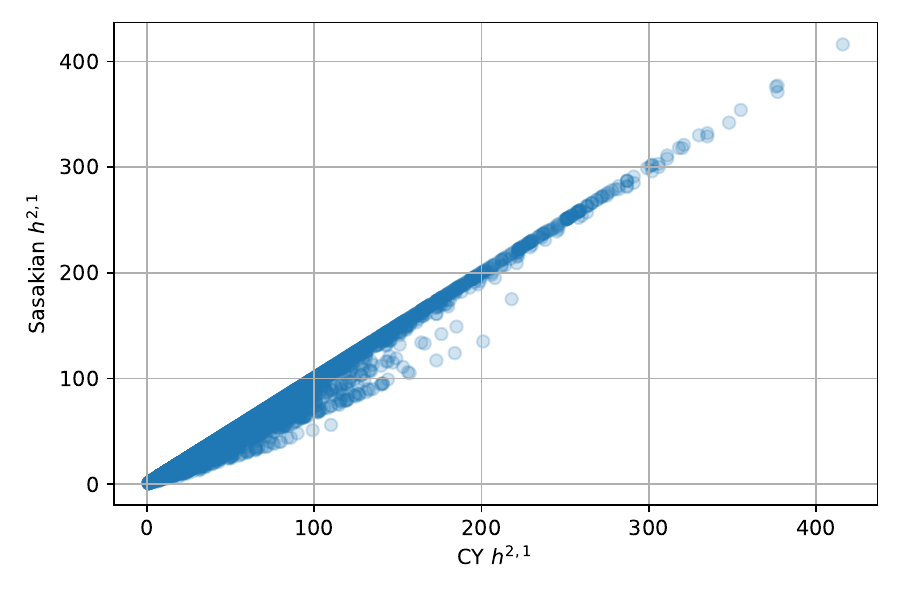}
    \caption{Comparison of the $h_S^{2,1}$ values for each Calabi-Yau link in the generated database to the $h^{2,1}$ value of the Calabi-Yau 3-fold used in its construction.}\label{fig:h21scatter}
\end{figure}

Further to the equivalent learning of $h_S^{2,1}$ as has been done for $h^{2,1}$ of the Calabi-Yau 3-folds used in the construction, one may also directly compare their values, as shown in Figure \ref{fig:h21scatter}.
As illustrated in the figure, an unexpected upper bound is set by the Calabi-Yau 3-fold's $h^{2,1}$ value, which is saturated for just over half of the 7555 cases. 
This lead to another novel conjecture about the link construction method
\newpage
\medskip
\noindent \textbf{Conjecture 2}: \textit{For Sasakian Hodge numbers $h_S^{p,q}$ of a 7-dimensional Calabi-Yau link, and Calabi-Yau Hodge numbers $h_{CY}^{p,q}$ of the respective Calabi-Yau 3-fold used to build the link, $h_{S}^{2,1} \leq h_{CY}^{2,1}$}.
\medskip

\noindent noting the analogous bound also technically holds for $1 = h_{S}^{3,0} \leq h_{CY}^{3,0} = 1$; as it may well for other yet uncomputed Sasakian Hodge numbers also.

\paragraph{Example:} For our running example of \eqref{eq:example}, permuting the weight system order and resampling the coefficients $a_i$ (ensuring the correct singularity structure is maintained) produces an equivalent Calabi-Yau 3-fold:
\begin{align}
    & \text{Weights}: \ [75, 22, 49, 29, 50]\,,\\
    & 0 = 48z_1^3 + 49z_1z_2z_3z_4z_5 + 6z_1z_5^3 + 71z_2^8z_3 + 35z_1^4z_4^3z_5 + 29z_2z_4^7 + 25z_3^4z_4\,.
\end{align}
Computing the respective link for this and then its equivalent topological invariants gives values: $\{h^{3,0},h^{2,1}\} = \{1,2\}$ and $\nu = 27$, matching the original values and exemplifying the corroboration of Conjecture 1.
Additionally, considering the $h^{2,1}$ value for the Calabi-Yau manifold in \eqref{eq:example} directly gives $h_{CY}^{2,1} = 2$, matching the $h_S^{2,1}$ value and corroborating Conjecture 2, satisfying the bound.

\paragraph{Outlook:} The work of \cite{Aggarwal:2023swe}, summarised here as an invited contribution to the DANGER proceedings, generated the largest database of Calabi-Yau links to date, identified members of new topological classes from the link construction, suggested the existence of a direct formula for Sasakian Hodge numbers from the ambient weight system information alone, raised conjectures about bounds on these link Sasakian Hodge numbers, and conjectured more broadly relations between their respective Milnor algebras (implying the database of computed invariants are truly exhaustive).

Future work looks to prove these raised conjectures, expand the applications of machine learning to Gr\"obner bases, and with the help of interpretable methods uncover the prospective formula for Calabi-Yau link Sasakian Hodge numbers from the weight information.

\section*{Acknowledgements}
The author wishes to thank Daattavya Aggarwal, Yang-Hui He, Elli Heyes, Henrique N. S\'a Earp, and Tom\'as S. R. Silva for collaboration on the work summarised here; and acknowledges support from Pierre Andurand over the course of this research.

\linespread{0.9}\selectfont
\addcontentsline{toc}{section}{References}
\bibliographystyle{utphys}
\bibliography{references}

\end{document}